\documentclass[twoside,english]{iopart}
\usepackage[T1]{fontenc}
\usepackage[latin9]{inputenc}
\usepackage{geometry}
\usepackage{float}
\usepackage{amstext}
\usepackage{amssymb}
\usepackage{graphicx}
\usepackage{esint}

\makeatletter

\providecommand{\tabularnewline}{\\}

\usepackage{iopams}
\usepackage{setstack}

\usepackage[numbers, sort&compress]{natbib}

\makeatother

\usepackage{babel}
\begin{document}

\title[A 3D aKMC study of dynamic solute-interface interaction]{{\Large A Three-Dimensional Atomistic Kinetic Monte Carlo Study of
Dynamic Solute-Interface Interaction}}

\author{{\normalsize A T Wicaksono$^{1,2}$, C W Sinclair$^{1,2}$ and M
Militzer}$^{1,2}$}

\address{\textup{$^{1}$The Centre for Metallurgical Process Engineering,
The University of British Columbia, Vancouver, Canada, V6T 1Z4 }}

\address{\textup{$^{2}$Department of Materials Engineering, The University
of British Columbia, Vancouver, Canada, V6T 1Z4}}

\ead{tegar@alumni.ubc.ca}
\begin{abstract}
A three-dimensional atomistic Kinetic Monte Carlo model was developed
and used to study the interaction between mobile solutes and a migrating
interface. While the model was developed with a simplified energetic
and topological description, it was also constructed to capture, in
the absence of solute, the Burke-Turnbull model for interface migration
and, in the presence of solutes, solute segregation to different types
of interface sites. After parameterizing the model, simulations were
performed to study the relationship between average interface velocity
and imposed driving pressure for varying solute concentration and
solute diffusivity. While the effect of solute concentration on solute
drag pressure was found to be consistent with classical solute drag
models, the effect of solute diffusivity was found to give a response
not captured by either continuum or previously reported two-dimensional
atomistic models. The dependence of maximum drag pressure on solute
diffusivity was observed and attributed to the coupling between the
structure of a migrating interface and the ability for solute to remain
segregated to the interface. 
\end{abstract}
\maketitle

\section{Introduction}

The dynamic interaction between solute atoms and migrating crystalline
interfaces is crucial for determining the microstructure of many materials
of technological relevance. For example, microalloying low-carbon
steels with Nb promotes the formation of fine-grained ferrite resulting
in an associated increase in strength and toughness \cite{Niobium}.
Niobium in solid solution drastically retards grain boundary migration
during recrystallization, and $\gamma/\alpha$-interface migration
during phase transformations \cite{austenite-nb-recryst}. Classically,
this phenomenon is described at steady state by the solute drag models
of L\"{u}cke-Detert \cite{LuckeDetert}, Cahn \cite{Cahn1960554},
L\"{u}cke-St\"{u}we \cite{Lucke19711087}, and Hillert \cite{Hillert1975,Hillert04}.
These continuum models predict the retardation of interface motion
due to a non-equilibrium solute distribution at the interface. This
distribution develops as a consequence of the competition between
the solute segregation and interface migration. As pointed out by
Hillert \cite{Hillert04}, the non-equilibrium solute profile leads
to a dissipation of a portion of the available driving force for interface
motion. 

While the concept of solute drag is broadly accepted, few attempts
have been made to directly evaluate the models against experimental
observations, e.g. by fitting experimental data using the models in
order to extract the model parameters \cite{Solutedragexperiments,Sinclairsolutedrag,Sinclairsolutedragcombined}.
One reason is that a direct comparison requires parameters that are
difficult to assess experimentally. For example, the models require
a knowledge of interface width, the spatial profile of solute-interface
binding energy and the trans-interface solute diffusivity. 

Recently, atomistic models have been used to examine the physical
origins of these parameters as well as to test some of the basic assumptions
of the continuum models \cite{MendelevSrolovitzRev,PFC}. Tackling
solute drag at the atomistic scale is challenging since the simulations
require diffusive time scales to resolve the rate of interface migration
in, for example, curvature-driven grain growth, recrystallization
and diffusive phase transformations. 

While molecular dynamic (MD) simulations have progressed to the point
where interface velocities at the upper end of those obtained experimentally
can be simulated \cite{PhysRevB.84.214102}, these velocities are
still too high to capture velocities where solute-interface interactions
are important \cite{MishinReview}. Using the recently developed phase
field crystal (PFC) technique that is capable of combining diffusive
time-scales and atomistic length-scales \cite{GreenwoodPFC}, an attempt
has been made to analyze solute drag problems in a non-ideal binary
system \cite{PFC}. Owing to computational overhead initial PFC simulations
of solute drag have focused on two-dimensional domains. In a first
approximation, the results reported from PFC simulations are consistent
with Cahn's solute drag model.

One of the most commonly employed simulation tools for tackling problems
that involve atomistic length-scales and diffusive time-scales is
atomistic kinetic Monte Carlo (aKMC). Mendelev et al \cite{MendelevSrolovitzRev,MendelevSrol-PhilMagA}
applied this technique to study the migration of a driven interface
in a two-dimensional, binary system where solute drag in a non-ideal
solid solution was examined. Departures from continuum models were
observed, notably an asymmetric effect between attractive and repulsive
solute-interface interaction on the drag pressure was reported \cite{MendelevSrolovitzRev}.
While these simulations provide valuable insights into possible atomic
scale contributions missed in the classic continuum solute drag models,
the limitation to two-dimensions and simplified energetics limit the
generality of the conclusions that can be drawn.

In the work presented here, atomistic kinetic Monte Carlo simulations
of solute drag have been extended to three dimensions. The paper is
organized as follows. First, the model is presented starting from
a description of the interface structure. The energetics and kinetics
of the system are described next. The simulation results are then
discussed and finally evaluated in the context of the continuum Cahn
model \cite{Cahn1960554}.

\section{Simulation Methodology}

\subsection{Geometry and energetics of a single component bicrystal}

The approach adopted here has been to focus on developing a model
that captures many of the basic features of a migrating crystaline
interface interacting with mobile solute atoms without capturing the
crystallographic detail of the interface itself. To do this, the simulation
box was constructed from a single body centered cubic (BCC) crystal
containing $N_{\textrm{X}}\times N_{\textrm{Y}}\times N_{\textrm{Z}}$
unit cells. The box height $N_{\textrm{Z}}$ is 200 unit cells, chosen
such that a steady state interface motion can be observed. The width
$N_{\textrm{X}}$ and length $N_{\textrm{Y}}$ were systematically
increased to determine the critical temperature for roughening transition
\cite{Olmsted20071161}. All subsequent simulations were conducted
in a system with $N_{\textrm{X}}=N_{\textrm{Y}}=120$ unit cells where
the finite-size effect, as indicated by the relative increase of interface
roughness at a given temperature with increasing system size, was
found to be no longer significant, i.e. a less than 10\% change. 

\begin{figure}[H]
\begin{centering}
\includegraphics[scale=0.5]{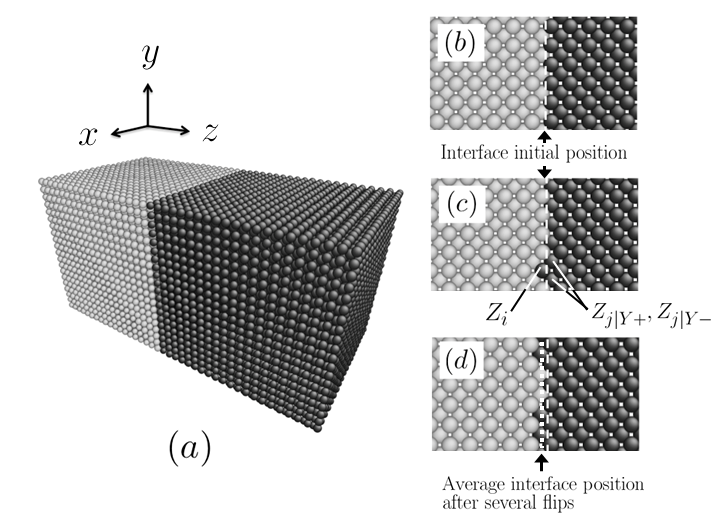}
\par\end{centering}

\caption{\label{fig:Interfaceview}(a) A typical simulation box where two crystals
of the same orientation are separated by a flat \{001\} interface,
(b) The initial position of the interface, (c) A flipping event occured
at the spin $i$ whose neighbours are $j$, (d) The average position
of the interface advances upon several flipping events of the spins
belonging to one of the grains.}
\end{figure}

In this construction, rather than being defined as the region separating
two crystals having different orientations, an interface was constructed
from a domain having a single crystallographic orientation by assigning
a \textquoteleft{}spin\textquoteright{} to each atom. The atoms in
one half of the bicrystal were assigned a \textquoteleft{}spin\textquoteright{}
of $+\frac{1}{2}$ (dark-coloured atoms in Figure \ref{fig:Interfaceview})
while those in the other half of the box were assigned a \textquoteleft{}spin\textquoteright{}
of $-\frac{1}{2}$ (light-coloured atoms in Figure \ref{fig:Interfaceview}).
Interfacial atoms are those who have at least one nearest neighbour
that belongs to the other grain. One can envision this as a coherent
interface separating two crystals having the same orientation and
lattice parameter, but not necessarily the same energy. Periodic boundary
conditions were applied to the simulation walls perpendicular to the
interface plane, while a toroidal boundary condition was applied to
the walls parallel to the interface plane \cite{MendelevSrol-PhilMagA}.
These boundary conditions ensure that the simulation box contains
only one interface plane. 

Coinciding with this simplified interfacial structure, a simplified
mechanism of interface migration was also adopted. Only solvent atoms
residing at the interface plane (Figure \ref{fig:Interfaceview}(b))
can flip their membership from their current grain to the adjacent
grain by switching their spin, effectively shifting the average position
of the interface, see Figure \ref{fig:Interfaceview}(d). The rate
of these flipping events for a given atom is assumed to depend on
its local environment. Without an energy bias between the two crystals,
the flipping events occur randomly to both the left and right. This
results in an interface that roughens but whose average position remains
the same. If, however, one of the crystals is assumed to have a higher
energy than the other, flipping will be biased in one direction causing
the interface to migrate.

Taking the energy of a system containing a flat interface as a reference,
the excess energy associated with a non-flat interface is assumed
to be \cite{CrystalGrowth},

\begin{equation}
E_{\textrm{pure}}=\gamma\underset{{\scriptstyle i=1}}{\overset{{\scriptstyle N_{\textrm{X}}N_{\textrm{Y}}}}{\sum}}\underset{{\scriptstyle j=1}}{\overset{{\scriptstyle 4}}{\sum}}\left(Z_{i}-Z_{j}\right)^{2}\label{eq:F11_F12}
\end{equation}

In this description, similar to the Discrete Gaussian Solid on Solid
(DGSOS) model \cite{RougheningSurface}, the excess energy of a rough
interface is characterized by a set of half-integral multiple of lattice
parameter $Z_{i}$ indicating the position along z-direction of an
interfacial atom of a given spin from an arbitary XY-plane of reference,
here taken as the initial interface position. The excess energy is
given by the sum of the square of the differences between the height
of each atom $Z_{i}$ and those of adjacent interfacial atoms $j$,
$Z_{j|\textrm{X}+},$ $Z_{j|\textrm{X}-},$ $Z_{j|\textrm{Y}+}$ and
$Z_{j|\textrm{Y}-}$, where $i$ and $j$ belong to the same type
of spin (see Figure \ref{fig:Interfaceview}(c)). The magnitude of
the energy penalty due to roughening is scaled by $\gamma$, which
can be thought of as an effective surface energy with units of energy
per unit area.

\subsection{Introducing solutes: energetics and interaction with interface}

From the single component bicrystal described above, binary alloys
were constructed having solute atoms residing in octahedral interstitial
sites of the BCC lattice. We have assumed identical behaviour for
solute in the two grains, with no solute-solute interactions.

\begin{figure}[H]
\begin{centering}
\includegraphics[scale=0.37]{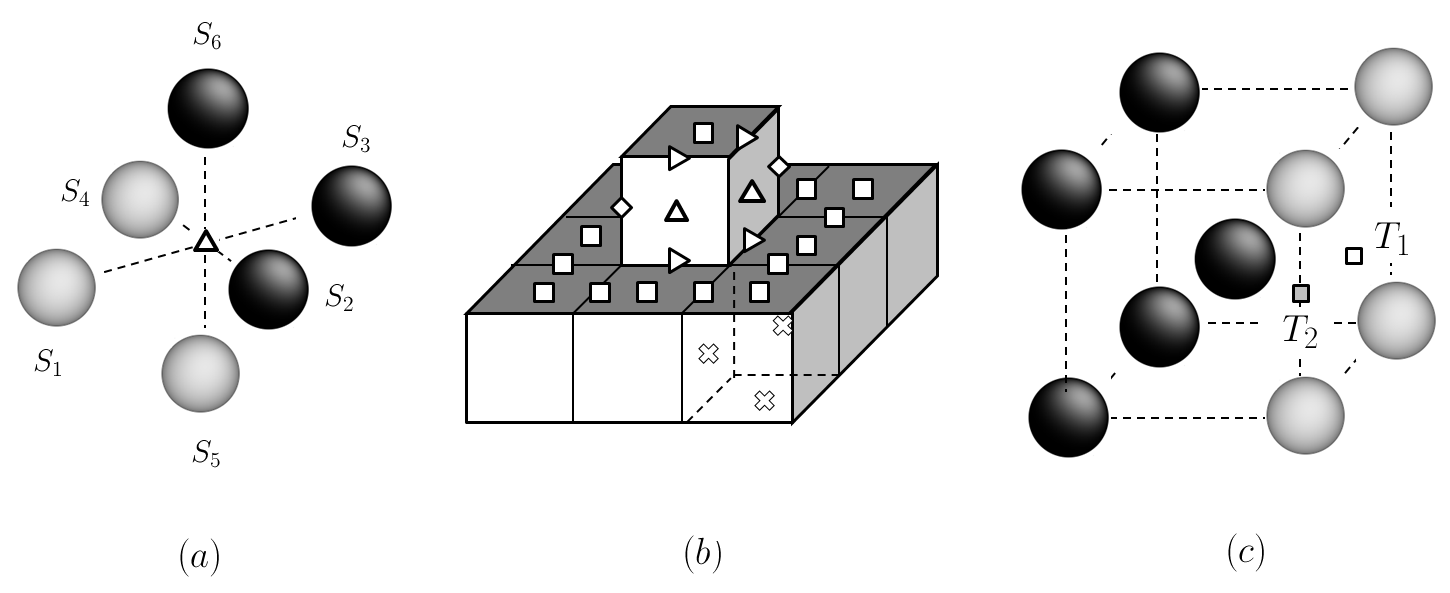}
\par\end{centering}

\caption{(a) An octahedral site in a BCC lattice surrounded by its six neighbouring
solvent atoms, (b) Different types of octahedral site as defined by
Equation (\ref{eq:spins}), see also Table \ref{tab:Types-of-interstitial},
(c) Two multiplicities of terrace site ($\omega=1.0$) found in a
flat interface.\label{fig:Octa-multiple}}
\end{figure}

The interaction between solute atoms and the interface were designed
such that the solute prefers to sit in positions where it is surrounded
by solvent atoms from both sides of the interface. An octahedral site
is surrounded by six solvent atoms, which we pair into three groups:
$(S_{1}:S_{3})$, $(S_{2}:S_{4})$ and $(S_{5}:S_{6})$, see Figure
\ref{fig:Octa-multiple}(a), the midpoint of each being the octahedral
site itself. The total energy of the system is modified when a solute
atom occupies a site that separates two solvent atoms belonging to
different grains. Using this convention a bulk octahedral site is
defined as one surrounded entirely by solvent atoms from the same
grain. The energy of non-bulk sites are scaled based on the parameter
$\omega$, where

\begin{eqnarray}
\omega & = & 0.5\left(|S_{1}-S_{3}|+|S_{2}-S_{4}|+2|S_{5}-S_{6}|\right)\nonumber \\
 &  & \textrm{with }S_{i}=\pm0.5\textrm{ (for left/right grain)}\label{eq:spins}
\end{eqnarray}

Based on Equation (\ref{eq:spins}), five distinct types of octahedral
sites can be identified. The factor two in the third term of Equation
(\ref{eq:spins}) is included to ensure that the two types of octahedral
site found in a perfectly flat interface ($T_{1}$ on the face center
and $T_{2}$ on the edge midpoint, see Figure \ref{fig:Octa-multiple}(c))
have $\omega=1$. The values of $\omega$ and their multiplicity for
the five types of sites are listed in Table \ref{tab:Types-of-interstitial}. 

\begin{table}[H]
\caption{Types of interstitial octahedral sites and their $\omega-$values,
see Equation (\ref{eq:spins}).\label{tab:Types-of-interstitial} }

\centering{}%
\begin{tabular}{|c|c|c|c|}
\hline 
Symbol & Site Type & $\omega-$values & Multiplicity\tabularnewline
\hline 
\hline 
$\times$ & Bulk & 0.0 & 2\tabularnewline
\hline 
$\vartriangleright$ & Ledge & 0.5 & 3\tabularnewline
\hline 
$\square$ & Terrace & 1.0 & 3\tabularnewline
\hline 
$\diamondsuit$ & Kink & 1.5 & 1\tabularnewline
\hline 
$\triangle$ & Island & 2.0 & 1\tabularnewline
\hline 
\end{tabular}
\end{table}

The binding energy of solute to an octahedral site is assumed to be
proportional to its $\omega$-value and is given by $E_{\textrm{bind}}=-\varepsilon_{\textrm{B}}\omega$.
Here, $\varepsilon_{\textrm{B}}$ is the binding energy to a site
in a flat interface, a positive value of which indicating an attractive
solute-interface interaction.

\subsection{Kinetics of interface migration and solute diffusion}

The system dynamics were simulated using a classic kinetic Monte Carlo
scheme where the kinetics are dictated by changes in the total energy
of the system before and after an event occurs \cite{Rautianen,KineticallyResolvedBarrier}.
In a bicrystal containing solute, the total system energy relative
to a solute-free system containing a flat interface is 

\begin{eqnarray}
E & = & E_{\textrm{pure}}+E_{\textrm{solute-interface}}\nonumber \\
 & = & \left(\gamma\underset{{\scriptstyle i=1}}{\overset{{\scriptstyle N_{\textrm{X}}N_{\textrm{Y}}}}{\sum}}\underset{{\scriptstyle j=1}}{\overset{{\scriptstyle 4}}{\sum}}\left(Z_{i}-Z_{j}\right)^{2}\right)+\left(-\varepsilon_{\textrm{B}}\cdot\underset{{\scriptstyle l=1}}{\overset{{\scriptstyle 6N_{\textrm{X}}N_{\textrm{Y}}N_{\textrm{Z}}}}{\sum}}\omega_{l}\phi_{l}\right)\label{eq:totalenergy}
\end{eqnarray}
where the summation index $l$ is over all octahedral sites and $\phi$
denotes the occupancy of site $l$, i.e. one if occupied and zero
otherwise. A fast-searching algorithm \cite{PhysRevE.51.R867} was
implemented along with the residence time method \cite{FichthornWeinberg}. 

The rates of two fundamental events have to be considered in this
model: the switching of interfacial atoms from one grain to the other
and solute diffusion. In the simplest case, the former is taken to
occur with a rate, 

\begin{equation}
\Gamma_{\textrm{int}}=\nu\cdot\exp\left(-\frac{Q_{\textrm{m}}+\Delta E/2}{kT}\right)\label{eq:rate_forwardmigration}
\end{equation}
where $\nu$ is the attempt frequency, $Q_{\textrm{m}}$ is the activation
barrier for interface migration and $\Delta E$ is the difference
between the total energy of the system before and after the event,
$k$ is the Boltzmann constant and $T$ is the absolute temperature
\cite{Rautianen,KineticallyResolvedBarrier}. The energy change $\Delta E$
in this case arises from changes of interfacial energy, via $\left(Z_{i}-Z_{j}\right)$
from the first term in Equation (\ref{eq:totalenergy}), as well as
changes in the number of occupied non-bulk octahedral sites due to
the interface motion away from segregated solutes, via $\omega_{l}$
from the second term in Equation (\ref{eq:totalenergy}). 

Under these conditions, the average position of the interface will
fluctuate around its initial average position. To drive the interface
in one direction, a difference between the energy of solvent atoms
belonging to the two grains must be imposed. This is done by raising
the energy of solvent atoms on one side of the interface by an amount
of $P\Omega$, $P$ being the driving pressure and $\Omega$ the atomic
volume. The rate of flipping in one direction is then still governed
by Equation (\ref{eq:rate_forwardmigration}) while the rate of flipping
in the other direction is given by

\begin{equation}
\Gamma_{\textrm{int-reverse}}=\nu\cdot\exp\left(-\frac{Q_{\textrm{m}}+P\Omega+\Delta E/2}{kT}\right)\label{eq:rate_reversemigration}
\end{equation}
Written in this way, the rate of interface migration for a pure system
obeys the classic Burke-Turnbull relationship \cite{BurkeTurnbull}.
This approximates to a linear dependence between $P$ and interface
velocity at low driving pressures, but more generally to a non-linear
relationship at high driving pressures where the velocity approaches
a limiting value. This contrasts with the interface migration model
in \cite{MendelevSrol-PhilMagA} where the velocity increases exponentially
with $P$.

The second rate that has to be captured is that associated with solute
diffusion. The rate of solute hops from one octahedral site to the
next is governed by, 

\begin{equation}
\Gamma_{\textrm{solute}}=\nu\cdot\exp\left(-\frac{Q_{\textrm{d}}+\Delta E/2}{kT}\right)\label{eq:rate_solutediffusion}
\end{equation}
where $Q_{\textrm{d}}$ is the activation barrier for bulk solute
diffusion and the energy change $\Delta E$ in this case comes only
from changes in the occupancy parameter $\phi$ from the second term
of Equation (\ref{eq:totalenergy}). Based on the above description,
the model has been parameterized using values that are consistent
with those expected in metallic alloys. Table \ref{tab:parameters}
summarizes the definition and the values of all fixed parameters chosen
for simulations. Parameters that will be varied during simulations
are the driving pressure $P,$ the activation barrier for bulk solute
diffusion $Q_{\textrm{d}}$, the absolute temperature $T$ and the
solute concentration $C_{0}$.

\begin{table}[H]
\caption{\label{tab:parameters}Basic simulation parameters.}

\noindent \centering{}%
\begin{tabular}{|c|l|c|}
\hline 
Parameter & Definition & Values\tabularnewline
\hline 
\hline 
$a$ & Lattice parameter & 0.3 nm\tabularnewline
\hline 
$\gamma$ & Surface energy & 52.8 mJ/m$^{2}$\tabularnewline
\hline 
$\varepsilon_{\textrm{B}}$ & Binding energy of solute to the interface & 0.17 eV\tabularnewline
\hline 
$\nu$ & Attempt frequency & 10$^{13}$ s$^{-1}$\tabularnewline
\hline 
$\Omega$ & Atomic volume & 2.7$\times$10$^{-29}$ m$^{3}$\tabularnewline
\hline 
$Q_{\textrm{m}}$ & Activation energy for interface migration & 0.1 eV\tabularnewline
\hline 
\end{tabular}
\end{table}

\section{Results}

\subsection{Interface migration in a solute-free bicrystal}

As a starting point for our investigations, simulations were performed
to establish the behaviour of the system in the absence of solute.
As has been previously reported for similar DGSOS models \cite{RougheningSurface},
the interface undergoes a roughening transition at a critical temperature
$T_{\textrm{c}}$. Identifying this transition is important since
the temperature at which the interface operates relative to $T_{\textrm{c}}$
determines its structure and thus its dynamics. To determine $T_{\textrm{c}}$,
a series of simulations were performed at different temperatures and
different system sizes with a fixed surface energy $\gamma$ = 52.8
mJ/m$^{2}$. The structure of the interface was monitored via the
time-average of its roughness $R$ \cite{Olmsted20071161}, 
\begin{equation}
R=\left(\frac{1}{N_{\textrm{X}}N_{\textrm{Y}}-1}\underset{{\scriptstyle i=1}}{\overset{{\scriptstyle N_{\textrm{X}}N_{\textrm{Y}}}}{\sum}}\left(Z_{i}-\overline{Z}\right)^{2}\right)^{1/2}\label{eq:stdv_roughness}
\end{equation}
 where $\overline{Z}$ is the average interface position.

Using the finite-size scaling method \cite{BinderTextbook} $T_{\textrm{c}}$
was found to be approximately $kT_{\textrm{c}}/a^{2}\gamma\sim2.42$,
corresponding to $T_{\textrm{c}}=833$ K for the surface energy and
lattice parameter in Table \ref{tab:parameters}. Below $T_{\textrm{c}}$,
the interface remains relatively flat, consisting predominantly of
terrace octahedral sites. 

\begin{figure}[H]
\begin{centering}
\includegraphics[scale=0.8]{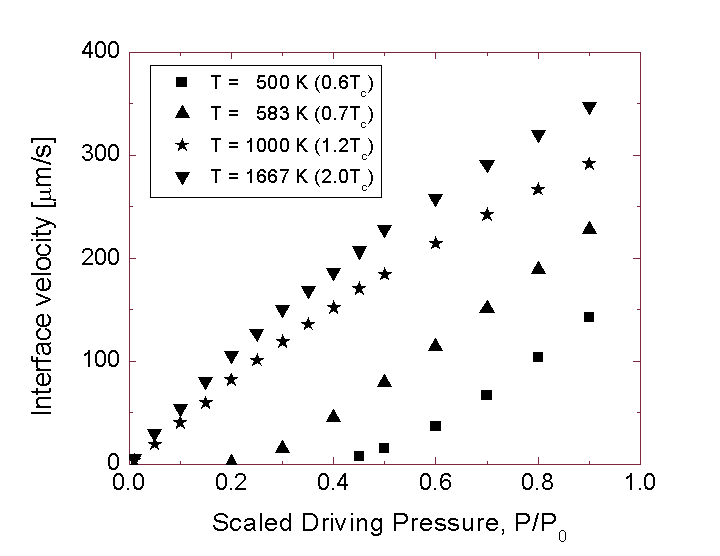}
\par\end{centering}

\caption{Dependence of average interface velocity on imposed driving pressure
$(P)$ for a solute-free system with $\gamma$ = 52.8 mJ/m$^{2}$.
The driving pressure is scaled by $P_{0}=kT/\Omega$.\label{fig:Pure_VvsP}}
\end{figure}

Figure \ref{fig:Pure_VvsP} illustrates the dependence of average
interface velocity on the imposed driving pressure $P$ for temperatures
above and below the critical temperature. The average interface velocity
was obtained at a given temperature by obtaining the slope of a linear
fit to the average interface position ($\overline{Z}$) as a function
of time. Each point on this plot represents the average of five different
simulations, the variation between runs being smaller than the size
of the symbols in this figure. 

At $T<T_{\textrm{c}}$, the interface motion is determined by a two-step
process of nucleation of an island, e.g. the atom $i$ in Figure \ref{fig:Interfaceview}(c),
followed by lateral propagation of ledges surrounding the nucleated
island. Owing to the high barrier for island nucleation relative to
$kT$, the rate of interface migration at low temperatures and low
driving pressures is negligibly small. In order to have a significant
motion of the interface, a critical driving pressure has to be applied,
its magnitude increasing with decreasing temperatures. 

At $T\geq T_{\textrm{c}}$, the barrier for island nucleation is reduced
relative to $kT$ and the interface migrates predominantly by spatially
uncorrelated island nucleations. The interface roughens and the velocity-driving
pressure relationship obeys the Burke-Turnbull model, where for sufficiently
small driving pressures the relationship is approximately linear \cite{BurkeTurnbull}. 

Based on the above results, all subsequent simulations were performed
at $T=1000\textrm{ K}>T_{\textrm{c}}$ so as to allow for a direct
comparison with predictions from continuum models, where a linear
velocity-driving pressure relation is typically assumed in the low
driving pressure limit.

\subsection{Interface migration in the presence of diffusing solutes}

In all simulations involving binary alloys, the system was populated
with a random distribution of solute. A driving pressure was next
imposed and the interface was observed to migrate. A transient regime
ensued until a steady-state distribution of solute segregated to the
interface was achieved. Upon reaching steady-state, the average interface
displacement was found to vary linearly with time, as in the case
of simulations of the single-component bicrystals described above.

\begin{figure}[H]
\begin{centering}
\includegraphics[scale=0.55]{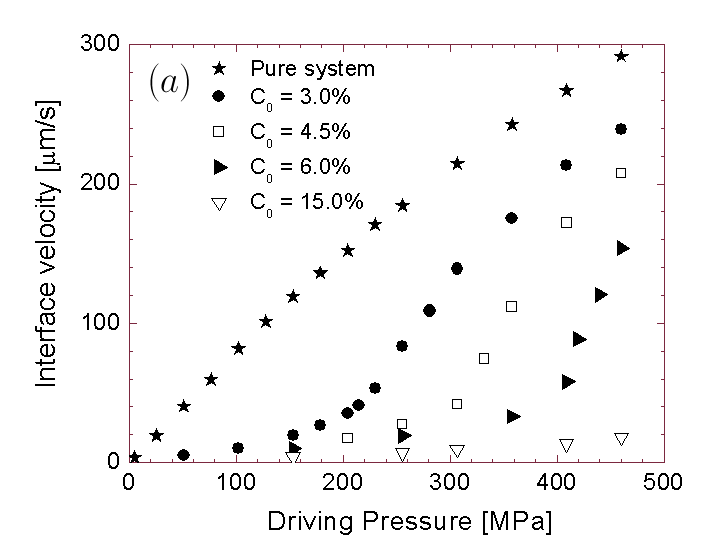} \includegraphics[scale=0.55]{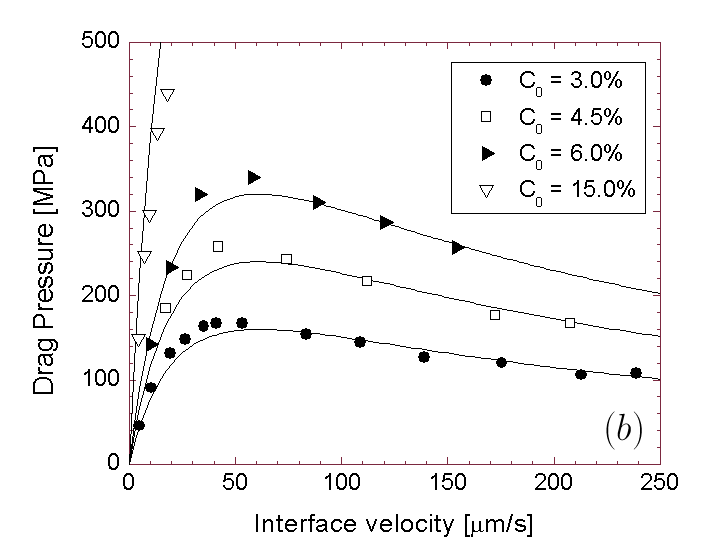}
\par\end{centering}

\caption{(a) Interface velocity as a function of imposed driving pressure for
systems containing different concentration of solute with $Q_{\textrm{d}}=0.27$
eV, (b) Drag pressure calculated and plotted versus velocity based
on data in (a). The solid curves were drawn using Equation (\ref{eq:modifiedCahn})
with $\alpha=$ 348 MPa($\mu$m/s)$^{-1}$ and $\beta=$ 1.63 $\times$10$^{-2}$
($\mu$m/s)$^{-1}$. \label{fig:varyconc}}
\end{figure}

Both solute concentration and solute diffusivity were varied to study
their effect on the rate of interface migration. The effect of solute
concentration on interface migration is shown in Figure \ref{fig:varyconc}
for a solute diffusion characterized by $Q_{\textrm{d}}=0.27$ eV.
At low driving pressures one observes that the interface velocity
decreases with solute concentration for a given driving pressure.
As the driving pressure is increased, the interface velocity of the
alloy system approaches that of the solute-free system, indicating
a breaking-away of the interface from its solute cloud. This qualitatively
obeys the classical description of solute drag from the continuum
models discussed earlier (see e.g. \cite{Cahn1960554}). 

Following Cahn \cite{Cahn1960554} and based on the results shown
in Figure \ref{fig:varyconc}(a), the drag pressure was computed as
the difference in driving pressure required to achieve the same velocity
in the solute-containing and solute-free system. For the solute-free
system the velocity-driving pressure relationship was fit to the Burke-Turnbull
model, $V=V_{T}\left(1-\exp\left(-P/P_{T}\right)\right)$ \cite{BurkeTurnbull}
where $\left(V_{T},P_{T}\right)$ were taken to be temperature-dependent
parameters with $V_{T}=553$ $\mu$m/s and $P_{T}=623$ MPa at 1000
K. A linear regime of velocity-driving pressure was observed for sufficiently
low driving pressures, i.e. $P\leq0.3P_{T}\approx200$ MPa. 

The calculated drag pressure $P_{\textrm{D}}$ is shown in Figure
\ref{fig:varyconc}(b) and compared with a continuum model that is
given by, 

\begin{equation}
P_{\textrm{D}}=\frac{\alpha C_{0}V}{\left(1+\beta V\right)^{2}}\label{eq:modifiedCahn}
\end{equation}
where $\alpha$ and $\beta$ are the model parameters. The model in
Equation (\ref{eq:modifiedCahn}) varies subtly from the one originally
proposed in \cite{Cahn1960554} and the origin of this expression
is outlined in \ref{sec:Modified-solute-drag}.

The aKMC results in Figure \ref{fig:varyconc}(b) are well described
by this expression using a single set of parameters $\alpha$ = 348
MPa ($\mu$m/s)$^{-1}$ and $\beta$ = 1.63 $\times$10$^{-2}$ ($\mu$m/s)$^{-1}$,
consistent with the fact that both are expected to be independent
of solute concentration \cite{Cahn1960554}. While Cahn provided the
explicit forms of $\alpha$ and $\beta$, both being a function of
the interface width, the diffusivity profile and the binding energy
profile, none of these are easily extracted from the present simulations.
Therefore, $\alpha$ and $\beta$ have been treated as adjustable
parameters. 

\begin{figure}[H]
\begin{centering}
\includegraphics[scale=0.63]{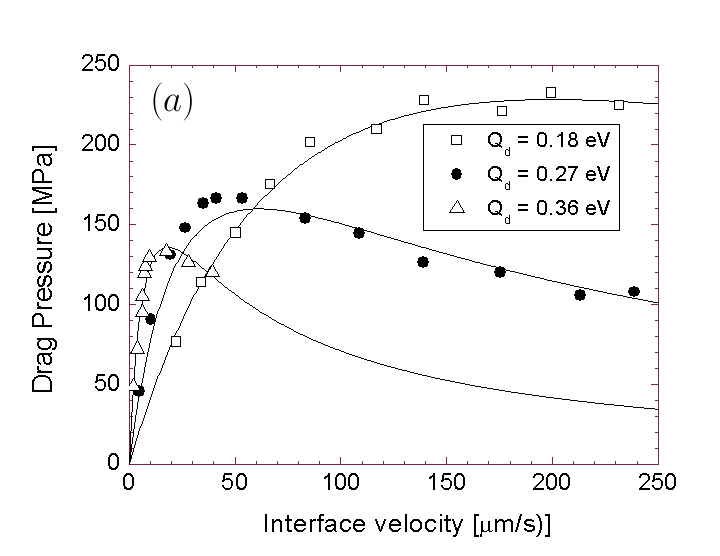} \includegraphics[scale=0.4]{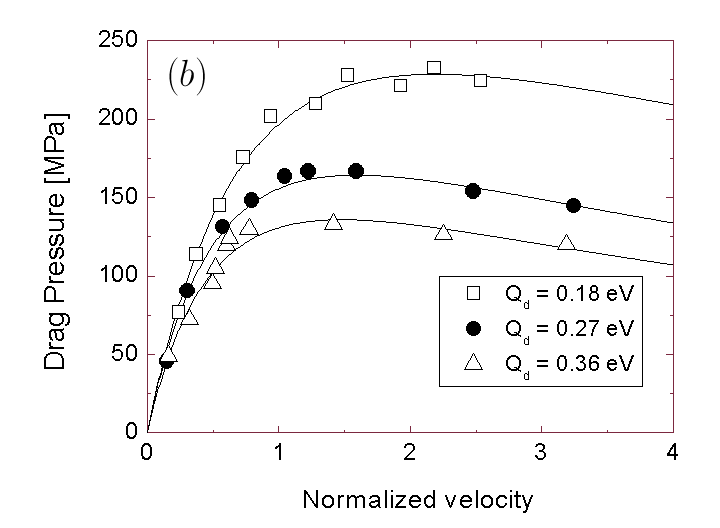}
\par\end{centering}

\caption{Drag pressure versus (a) velocity, and (b) normalized velocity for
systems containing 3 at\% solute and different solute diffusivities.
The normalized velocity is the velocity multiplied by the ratio of
lattice parameter and diffusivity. Solid lines indicate the empirical
fit between simulation results and empirical model, Equation (\ref{eq:modifiedCahn}).
\label{fig:VaryDiffus}}
\end{figure}

While the aKMC results and the Cahn's model agree when the role of
solute concentration is considered, the same cannot be said when the
effect of solute diffusivity on solute drag is examined. Figure \ref{fig:VaryDiffus}(a)
illustrates the velocity-driving pressure and drag pressure-velocity
relationships for systems having $C_{0}=$ 3.0 at$\%$ and different
solute diffusivities. Continuum solute drag models predict that these
curves would collapse when the interface velocity is scaled by the
solute diffusivity and a characteristic length-scale, typically taken
to be the interface width \cite{Cahn1960554}. Scaling the data in
this way (Figure \ref{fig:VaryDiffus}(b)) does not appear to bring
the normalized velocity at peak drag pressure into coincidence. The
continuum models would also predict that the peak drag pressure should
be independent of the solute diffusivity. Instead, Figure \ref{fig:VaryDiffus}(b)
reveals that the peak drag pressure increases with solute diffusivity.

\begin{figure}[H]
\begin{centering}
\includegraphics[scale=0.5]{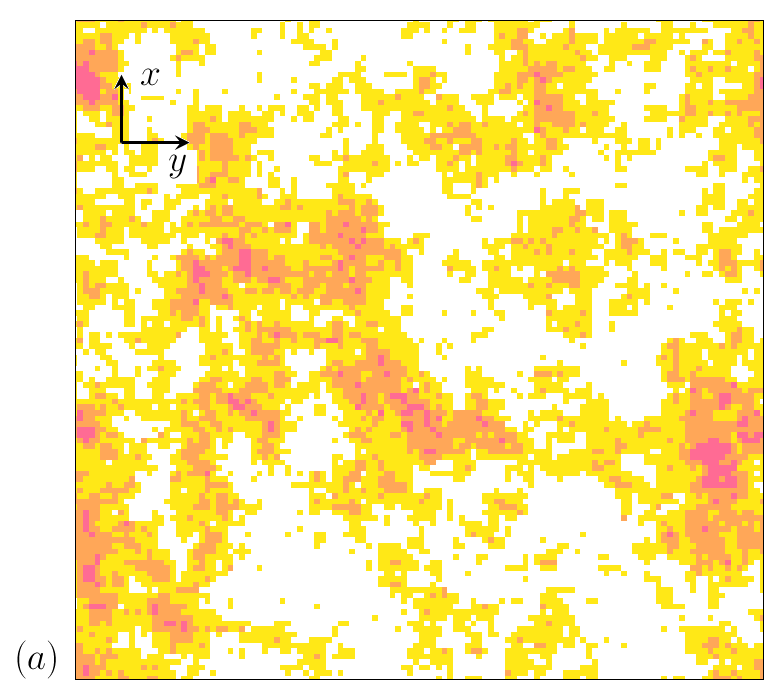}
\par\end{centering}

\begin{centering}
\includegraphics[scale=0.5]{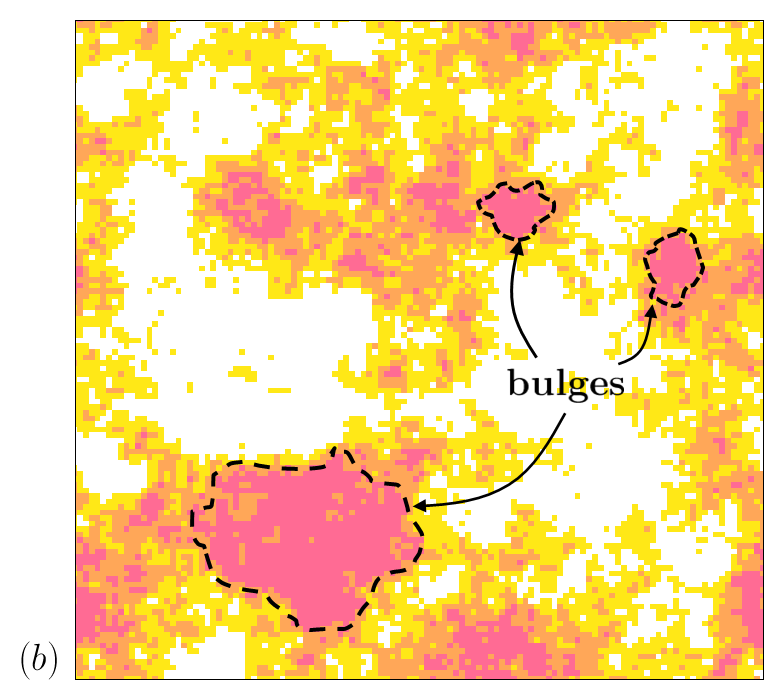}
\par\end{centering}

\caption{Snapshots of interface cross-sectional view during its steady-state
migration at the velocity corresponding to maximum drag pressure and
interacting with (a) slow-diffusing solute and (b) fast-diffusing
solute. Yellow pixels indicate interfacial solvent atoms $i$ whose
$\textrm{z}-$position is equal to the average position of the interface
($Z_{i}=\overline{Z}$). White, orange and salmon pixels indicate
$Z_{i}=\overline{Z}-a$, $Z_{i}=\overline{Z}+a$ and $Z_{i}>\overline{Z}+a$,
respectively. Regions corresponding to bulges in (b) are highlighted.
\label{fig:snapshots}}
\end{figure}

The deviation from classical continuum model can be attributed to
the fluctuation of interface topology during the course of its migration.
Observing the structure of the interface at the velocity corresponding
to the peak drag pressure reveals significant differences between
systems containing slow diffusing and fast diffusing solute. Figure
\ref{fig:snapshots} illustrates snapshots of the interface plane
under these conditions, particularly highlighting the roughness of
the interface plane. While the interfaces exhibit some similarities,
one can see that the interface interacting with fast diffusing solute
exhibits a higher degree of spatial correlation in the interface \textquoteleft{}height\textquoteright{}
leading to local bulging in several locations. Such bulges require
the coordinated motion of a large number of neighbouring interface
segments, suggesting a correlation of the local behaviour of interfacial
atoms. 

\begin{figure}[H]
\begin{centering}
\includegraphics[scale=0.4]{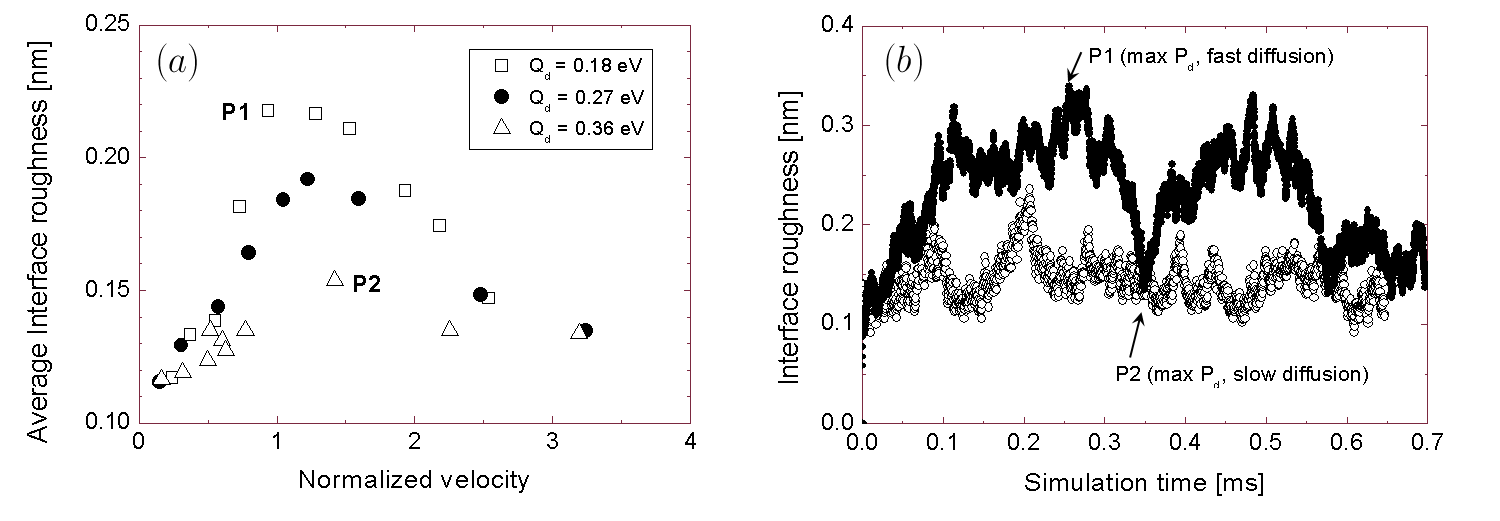}
\par\end{centering}

\caption{(a) The time-average interface roughness (Equation (\ref{eq:stdv_roughness}))
plotted against the normalized interface velocity, i.e. velocity$\times$
flat interface width/bulk diffusivity, for varying solute diffusivity;
(b) A time-resolved trace of interface roughness corresponding to
the peak drag pressure for interface interacting with fast-diffusing
(filled circles) and slow-diffusing solutes (open circles), i.e. point
P1 and P2 in (a). \label{fig:stdv_widthVaryDiffuse}}
\end{figure}

As illustrated in Figure \ref{fig:stdv_widthVaryDiffuse}(a), this
behaviour can be further quantified by examining the interface roughness
$R$ from each data point in Figure \ref{fig:VaryDiffus}(b). 

At low and high interface velocities, the interface remains relatively
flat irrespective of solute diffusivity, its roughness being similar
to that of a non-driven interface. In the case of systems containing
slow diffusing solutes, the interface roughness increases by less
than half its stationary value as the velocity approaches its value
at the peak drag pressure. In the case of fast-diffusing solute, the
interface roughness nearly doubles relative to its stationary value.
A direct consequence of the interface motion in the presence of fast
diffusing solute is a higher rate of energy dissipation, resulting
in a higher peak drag pressure. 

While this provides an explanation for the observed increase of maximum
drag pressure with increasing solute diffusivity, it does not explain
the relationship between solute diffusivity and interface topology.
The dependence of interface topology on solute diffusivity can be
explained from the perspective of the rate of solute hopping into
and away from a migrating interface. 

For a detailed discussion about this phenomenon, one may start by
considering a flat interface containing a steady-state distribution
of segregated solute. At some location along the flat interface plane,
a small interface segment will advance into the adjacent crystal due
to the applied driving pressure. The next atomistic event depends
sensitively on the diffusivity of the solute atoms surrounding this
advanced interface segment. 

At low diffusivity (high $Q_{\textrm{d}}$), the rate of solute hopping
is predominantly determined by the activation barrier $Q_{\textrm{d}}$,
the rate of hopping to and from the interface being approximately
the same, i.e. $\Gamma_{\textrm{bulk}\rightarrow\textrm{int}}^{\textrm{sol}}\approx\Gamma_{\textrm{int}\rightarrow\textrm{bulk}}^{\textrm{sol}}=\nu\exp\left(-Q_{\textrm{d}}/kT\right)$
since $Q_{\textrm{d}}$ is relatively large compared to $\varepsilon_{\textrm{B}}$.
The next event in the simulation is also likely to be the advance
of an adjacent interface segment. This is due to the imposed driving
pressure that favours this event as well as the fact that the system
energy can be lowered by preferentially having the neighbouring segments
to advance. This operation is expected to repeat along the interface
plane to reduce the total interface energy (Equation (\ref{eq:totalenergy})),
occuring more frequently than solute atoms diffusing into the sites
available in the advanced interface segments. The distribution of
interface height is thus expected to be largely uncorrelated as seen
in Figure \ref{fig:snapshots}(a). Moreover, it is unlikely that an
advanced interface segment will further proceed before the neighbouring
segments catch up because of the increased roughening penalty, see
Equation (\ref{eq:F11_F12}).

If the solute diffusivity is very high (low $Q_{\textrm{d}}$), however,
then immediately after the interface segment advances the solute atom
left behind has a high probability to jump back to a location in the
(now advanced) interface since the barrier for jumps into the interface
is significantly lowered by the binding energy of the solute to the
interface, $\Gamma_{\textrm{bulk}\rightarrow\textrm{int}}^{\textrm{sol}}\approx\nu\exp\left(-(Q_{\textrm{d}}-0.5\varepsilon_{\textrm{B}})/kT\right)$.
The rates of solute jumping out of the interface, on the other hand,
are low owing again to the binding energy of solute to the interface,
i.e. $\Gamma_{\textrm{int}\rightarrow\textrm{bulk}}^{\textrm{sol}}\approx\nu\exp\left(-(Q_{\textrm{d}}+0.5\varepsilon_{\textrm{B}})/kT\right)$.
Segregation of solute to a bulge on the interface will result in a
reduction in the total system energy, partially compensating for the
increased energy due to the larger total area of the interface. In
this situation, the next event can include a further extension of
the already advanced interface under the imposed driving pressure
since the energy penalty associated with this event is partially compensated
by the segregated solute. This is evident in Figure \ref{fig:snapshots}(b)
where the height distribution appears more spatially correlated compared
to Figure \ref{fig:snapshots}(a). The advance of a single interfacial
atom by more than two atomic positions ahead of its neighbours in
the same grain has a low probability owing to the rapid increase in
interfacial energy (Equation \ref{eq:F11_F12}). When a bulge extends
beyond this point, its growth will stall and the interface will flatten
so as to reduce the total energy of the system. Based on this explanation
and as observed in simulations (Figure \ref{fig:stdv_widthVaryDiffuse}(b)),
bulges will appear, grow then slow down and finally disappear as the
whole interface advances with new bulges appearing at other locations
on the interface plane. 

These results reveal a phenomenon that neither two-dimensional simulations
\cite{MendelevSrolovitzRev,PFC} nor continuum models have previously
revealed \cite{Cahn1960554}. The spatial degree of freedom available
for the structural fluctuation of a two-dimensional interface allow
for a more complex interfacial topography than the topographies reported
in two dimensional systems. While this leads to a non-classical dependence
of the peak drag pressure on solute diffusivity, it has to be emphasized
that, otherwise, the results presented here semi-quantitatively follow
the predictions of the classical solute drag models \cite{LuckeDetert,Cahn1960554,Lucke19711087}.
In some regards it is surprising that an effect of the interface structure,
arising explicitly from the different types of sites for solute at
the interface, was not found to play a significant role in results
even though the population of these different types of sites changes
depending on the topology (e.g. roughness) of the interface. 

Based on the ability of the Cahn's solute drag model to capture the
trends in terms of velocity-driving pressure, it is concluded that
the use of an effective binding energy is sufficient to describe the
solute drag effect observed here. A natural question that arises from
this is how best to compute such an effective binding energy profile
based on a discrete atomistic model. Similarly, quantities such a
trans-interface diffusivity and interface width do not have natural
analogues in atomistic simulations \cite{PFC,MDsolutedrag}. Given
that the parameters $\alpha$ and $\beta$ in the Cahn model depend
on these quantities, it is not possible to make a fully quantitative
comparison with atomistic simulations. Here, for example, $\alpha$
and $\beta$ have had to be treated as adjustable parameters. These
simulation results, however, may be used to develop useful ways of
finding effective parameters, e.g. the interface width, when modelling
experimental data. 

Finally, one must recognize the simplicity of the energetic and topological
descriptions of the solute-interface interactions employed here. These
descriptions limit the possible interface configurations being simulated.
Further simulations using more realistic interface/solute topology
and interactions need to be investigated to test the generality of
these findings and to extend the results to less idealized alloy systems.

\section{Summary}

An atomistic Kinetic Monte Carlo model has been developed having a
simplified description of system energy and interface topology. Using
this model the effect of diffusing solute on a migrating interface
has been investigated. The results are in semi-quantitative agreement
with the predictions of continuum models, though a departure from
these classic models was observed when the effect of solute diffusivity
on interface migration was examined. Unlike the classic models which
predict the peak drag pressure to be independent of solute diffusivity,
the simulations reported here showed an increasing peak drag pressure
with solute diffusivity. It was shown that this effect arises from
a change in the roughness of the interface at the peak drag pressure
depending on the solute diffusivity. Fast diffusing solutes allow
for local bulges to form and then disappear on the interface plane
during the interface migration. The formation of these bulges results
in a more tortuous path for interface migration and therefore a higher
rate of energy dissipation during migration. A fully quantitative
comparison with Cahn's solute drag model is limited by the fact that
the model parameters, such as binding energy profile, trans-interface
diffusivity and interface width, do not have one-to-one correspondence
in the atomistic simulations.

\ack{}{}

The authors thank the Natural Sciences and Engineering Research Council
of Canada (NSERC) for financial support. Valuable discussions with
Michael Greenwood and Michel Perez are gratefully acknowledged.

\appendix

\section{\label{sec:Modified-solute-drag}Modified solute drag model}

Cahn's solute drag model \cite{Cahn1960554} was developed by coupling
a steady-state diffusion equation, 

\begin{equation}
C^{\prime\prime}+\left(\varepsilon_{\textrm{B}}^{\prime}+\frac{D^{\prime}}{D}+\frac{V}{D}\right)C^{\prime}+\left(\frac{D^{\prime}}{D}\varepsilon_{\textrm{B}}^{\prime}+\varepsilon_{\textrm{B}}^{\prime\prime}\right)C=0\label{eq:Cahndiffusion}
\end{equation}
whose solution $C(x)$ requires assumptions about the spatial profile
of solute diffusivity $D(x)$ and the interfacial binding energy $\varepsilon_{\textrm{B}}(x)$,
to an expression for the drag pressure based on the force exerted
by each solute atom on the interface,

\begin{equation}
P_{\textrm{D}}=-\frac{1}{\Omega}\int_{-\infty}^{+\infty}\left(C(x)-C_{0}\right)\varepsilon_{\textrm{B}}^{\prime}dx\label{eq:dragpressure}
\end{equation}

While this coupled system of equations can be solved analytically
in the limit of low or high velocity, a completely general, closed
form expression for the drag pressure is not available. Cahn proposed
an expression to approximately merge responses obtained in the high
and low velocity limit. Cahn's proposed expression was \cite{Cahn1960554} 

\begin{equation}
P_{\textrm{D}}=\frac{\alpha C_{0}V}{1+\left(\beta V\right)^{2}}\label{eq:originalCahn}
\end{equation}

\begin{figure}[H]
\begin{centering}
\includegraphics[scale=0.75]{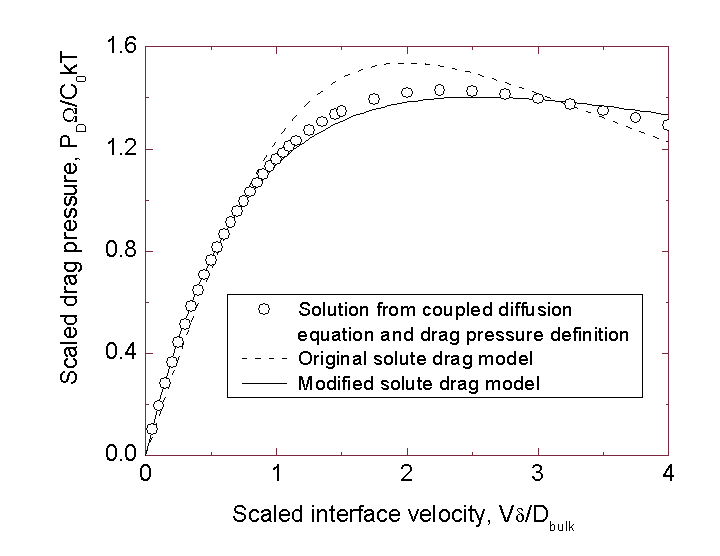}
\par\end{centering}

\caption{The drag pressure-velocity trend from solving coupled Equations (\ref{eq:Cahndiffusion})
and (\ref{eq:dragpressure}) that employ a non-uniform diffusivity
profile is fitted using Cahn's original model, Equation (\ref{eq:originalCahn}),
and the model proposed in this work, Equation (\ref{eq:modifiedCahn}).
Both the diffusivity and the binding energy spatial profile are triangular
where $D\left(x\leq-\frac{\delta}{2},\, x\geq+\frac{\delta}{2}\right)=D_{\textrm{bulk}},$
$D(x=0)=D_{\textrm{interface}}=4D_{\textrm{bulk}},$ $\varepsilon_{\textrm{B}}\left(x\leq-\frac{\delta}{2},\, x\geq+\frac{\delta}{2}\right)=0$
and $\varepsilon_{\textrm{B}}(x=0)=-2kT$.\label{fig:appendix}}
\end{figure}

When Equation (\ref{eq:Cahndiffusion}) assumes a constant diffusivity
and a triangular binding energy profile, explicit forms of $\alpha$
and $\beta$ were available \cite{Cahn1960554}. In this case, the
original model (Equation (\ref{eq:originalCahn})) is shown to fit
well the drag pressure obtained from numerically solving coupled Equations
(\ref{eq:Cahndiffusion}) and (\ref{eq:dragpressure}). If a non-uniform
diffusivity profile is assumed, e.g. the case where the trans-interface
diffusivity is different from the bulk diffusivity, the original model
(Equation (\ref{eq:originalCahn})) does not fit the numerical solution
as well, see the dashed curve in Figure \ref{fig:appendix}. A similar
trend is also observed when the original model (Equation (\ref{eq:originalCahn}))
is used to fit the numerical solutions to the modified diffusion equation
similar to Equation (\ref{eq:Cahndiffusion}) that includes site-saturation
correction \cite{Lucke19711087}. The parameters $(\alpha,\beta)$
used to draw the dashed curve in Figure \ref{fig:appendix} were calculated
from the slope and the ordinate intercept of linear regression $V^{2}$
versus $V/P_{\textrm{D}}$ where $(V,P_{\textrm{D}})$ are the points
obtained from numerically solving coupled Equations (\ref{eq:Cahndiffusion})
and (\ref{eq:dragpressure}). 

The models's inability to fit the data points relies on the fact that
the regression form mentioned earlier showed a parabolic instead of
linear trend. A slight modification to the model, Equation (\ref{eq:modifiedCahn}),
was proposed and it was found that this modified model fits the numerical
solution reasonably well, see the solid curve in Figure \ref{fig:appendix}.
Consistent with the fact that the present simulations involve a diffusivity
into and within the interface that are affected by the solute-interface
binding energy, it is found that the modified model also gives a better
match to the simulation results compared to the original model above.

\bibliographystyle{iopart-num}
\addcontentsline{toc}{section}{\refname}\bibliography{paper}

\providecommand{\newblock}{}
\begin{thebibliography}{10}
\expandafter\ifx\csname url\endcsname\relax
  \def\url#1{{\tt #1}}\fi
\expandafter\ifx\csname urlprefix\endcsname\relax\def\urlprefix{URL }\fi
\providecommand{\eprint}[2][]{\url{#2}}

\bibitem{Niobium}
Deardo A~J 2003 {\em Int. Mater. Rev.\/} {\bf 48} 371--402

\bibitem{austenite-nb-recryst}
Cuddy L~J 1982 {\em Thermomechanical Processing of Microalloyed Austenite\/}
  (TMS-AIME)

\bibitem{LuckeDetert}
L\"{u}cke K and Detert K 1957 {\em Acta Metall.\/} {\bf 5} 628--637

\bibitem{Cahn1960554}
Cahn J~W 1962 {\em Acta Metall.\/} {\bf 10} 789--798

\bibitem{Lucke19711087}
L\"{u}cke K and St\"{u}we H~P 1971 {\em Acta Metall.\/} {\bf 19} 1087--1099

\bibitem{Hillert1975}
Hillert M 1975 {\em Met. Trans. A\/} {\bf 6} 5--19

\bibitem{Hillert04}
Hillert M 2004 {\em Acta Mater.\/} {\bf 52} 5289--5293

\bibitem{Solutedragexperiments}
Zurob H~S, Zhu G, Subramanian S~V, Purdy G~R, Hutchinson C~R and Br\'{e}chet Y
  2005 {\em ISIJ International\/} {\bf 45} 713--722

\bibitem{Sinclairsolutedrag}
Sinclair C~W, Hutchinson C~R and Br\'{e}chet Y 2007 {\em Metall. Mater. Trans.
  A\/} {\bf 38} 821--830

\bibitem{Sinclairsolutedragcombined}
Hutchinson C~R, Zurob H~S, Sinclair C~W and Br\'{e}chet Y 2008 {\em Scr.
  Mater.\/} {\bf 59} 635--637

\bibitem{MendelevSrolovitzRev}
Mendelev M~I and Srolovitz D~J 2002 {\em Modell. Simul. Mater. Sci. Eng.\/}
  {\bf 10} R79--R109

\bibitem{PFC}
Greenwood M, Sinclair C~W and Militzer M 2012 {\em Acta Mater.\/} {\bf 60}
  5752--5761

\bibitem{PhysRevB.84.214102}
Deng C and Schuh C~A 2011 {\em Phys. Rev. B\/} {\bf 84} 214102

\bibitem{MishinReview}
Mishin Y, Asta M and Li J 2010 {\em Acta Mater.\/} {\bf 58} 1117--1151

\bibitem{GreenwoodPFC}
Greenwood M, Rottler J and Provatas N 2011 {\em Phys. Rev. E\/} {\bf 83} 031601

\bibitem{MendelevSrol-PhilMagA}
Mendelev M~I, Srolovitz D~J and E W 2001 {\em Phil. Mag. A\/} {\bf 81}
  2243--2269

\bibitem{Olmsted20071161}
Olmsted D~L, Foiles S~M and Holm E~A 2007 {\em Scr. Mater.\/} {\bf 57}
  1161--1164

\bibitem{CrystalGrowth}
Weeks J~D and Gilmer G~H 1979 {\em Adv. Chem. Phys.\/} {\bf 40} 157--228

\bibitem{RougheningSurface}
Lapujoulade J 1994 {\em Surf. Sci. Rep.\/} {\bf 20} 191--249

\bibitem{Rautianen}
Rautiainen T~T and Sutton A~P 1999 {\em Phys. Rev. B\/} {\bf 59} 13681--13692

\bibitem{KineticallyResolvedBarrier}
Van{ }der{ }Ven A, Ceder G, Asta M and Tepesch P~D 2001 {\em Phys. Rev. B.\/}
  {\bf 64} 184307

\bibitem{PhysRevE.51.R867}
Blue J~L, Beichl I and Sullivan F 1995 {\em Phys. Rev. E\/} {\bf 51} R867--R868

\bibitem{FichthornWeinberg}
Fichthorn K~A and Weinberg W~H 1991 {\em J. Chem. Phys.\/} {\bf 95} 1090--1096

\bibitem{BurkeTurnbull}
Burke J and Turnbull D 1952 {\em Prog. Met. Phys.\/} {\bf 3} 220--292

\bibitem{BinderTextbook}
Landau D~P and Binder K 2005 {\em A Guide to Monte Carlo Simulations in
  Statistical Physics, 2nd edition\/} (Cambridge University Press)

\bibitem{MDsolutedrag}
Mendelev M~I, Srolovitz D~J, Ackland G~J and Han S 2005 {\em J. Mater. Res.\/}
  {\bf 20} 208--218

\end{thebibliography}

\end{document}